\documentclass[aps,twocolumn,showpacs,prl,superscriptaddress]{revtex4}

\usepackage{dcolumn}
\usepackage{amsmath}
\usepackage{epsf}
\usepackage{graphicx}

\begin{document}

\title{Effect of Coulomb interaction on the spin-galvanic mode in a two dimensional electron gas with Rashba spin-orbit interaction}

\author{Ya.\ B. Bazaliy}
\affiliation{IBM Almaden Research Center, 650 Harry Road, San Jose, CA 95120}

\author{B. V. Bazaliy}
\affiliation{Applied Mathematics Institute, National Academy of Science of Ukraine, 70 R. Luxemburg, Donetsk, Ukraine}

\author{G. G\"untherodt}
\affiliation{II. Physikalisches Institut, Rheinisch-Westf\"alische Technische Hochschule Aachen, 52056 Aachen, Germany}

\author{S. S. P. Parkin}
\affiliation{IBM Almaden Research Center, 650 Harry Road, San Jose, CA 95120}

\date{\today}

\begin{abstract}
Recently a new propagating mode of coupled charge and spin oscillations was predicted in a two dimensional electron gas with a sufficiently strong Rashba interaction. We show that Coulomb interactions  qualitatively modifies the spectrum and increases the characteristic wavelength of this mode by orders of magnitude, but does not suppress it. An absorption experiment that can conclusively detect the presence or absence of such a propagating mode is proposed.
\end{abstract}



\maketitle

In the field of spintronics, semiconductor devices utilizing the spin degree of freedom enjoy special attention due to prospects of their ready integration with existing solid-state technologies \cite{semiconductor_spintronics_book}. In addition, the comparatively strong spin-orbit interaction in semiconducting materials allows for the possibility of spin manipulation by electric fields. This ameliorates the challenge of creating local and variable magnetic fields in the nanoscale regime to otherwise operate arrays of spintronic devices. The study of the interplay of spin and charge motion in semiconductors (see review \onlinecite{dyakonov_spintronics}) dates back to research on optical spin orientation~\cite{optical_orientation}. Subsequently, spin-galvanic effects were discovered, in which uniform spin polarization created by optical orientation or other techniques generates electric currents (see review \onlinecite{ganichev_review}). An inverse effect of inducing non-equilibrium spin density by current was suggested theoretically \cite{aronov_geller,edelstein_1990} and studied by several groups \cite{kalevich_korenev,awshalom_nature2003,silovAPL2004,awshcalom_nature_2005}. The spin-Hall effect, predicted in Ref.~\onlinecite{dyakonov_perel_spin_hall}, is another phenomenon in which electric current creates a spin response by generating a pure spin current in an orthohonal direction. A considerable discussion (see review \onlinecite{rashba_spintronic_review_2005}) of the nature of such a spin current is underway following two recent puclications~\cite{MurakamiScience2003,SinovaPRL2004} on the subject.

In this paper we consider a specific case of a Rashba two-dimensional electron gas (2DEG) system in the diffusive regime. With regard to potential applications, this regime is experimentally much less challenging than the ballistic regime, required in many other spintronic proposals. Moreover, in the 2DEG the spin-galvanic and spin-Hall effects are clearly distinguishable by the spin direction involved: the former is relevant for the in-plane spin density, while the latter creates a current of out-of-plane spin. When an alternating current is passed through a 2DEG, additional features may appear. Recently, a new propagating mode of coupled spin and charge oscillations was predicted in 2DEG structures with a sufficiently strong Rashba interaction~\cite{Bernevig}. A low frequency mode, existing in the diffusive regime, is of major potential importance in semiconductor physics. However, the derivation in Ref.~\onlinecite{Bernevig} treated electrons as neutral fermions. Such a treatment can be justified in the stationary cases~\cite{BurkovPRB2004,MishchenkoPRL2004}, but not in the case of the time dependent electron $\delta n(r,t)$ and associated charge $\rho = e \delta n$ density deviations in the propagating mode. In bulk conductors the presence of the Coulomb interaction prevents the existence of low frequency modes by imposing the quasi-neutrality condition, $\rho = 0$. An important exception are modes propagating in large external magnetic fields~\cite{theory_of_metals}, e.g., helicons. Since screening in two dimensions is weaker than in three dimensions, the quasi-neutrality condition is relaxed. But the uncompensated charges will create an induced electric field $E_{ind}$ which, in turn, will modify the motion of the system. A proper self-consistent account of the induced electric fields in the proposed mode is the main goal of the present paper.

Following Refs.~\onlinecite{Bernevig,BurkovPRB2004,MishchenkoPRL2004} we consider a 2DEG with the single-electron Hamiltonian
$$
H = \frac{p^2}{2m} + \alpha \hat z [\vec\sigma \times {\bf p}]
$$
Here $m$ is the effective mass, $\alpha$ is the Rashba interaction constant, $\hat z$ is a unit vector perpendicular to the plane of the 2DEG, and $\vec\sigma$ is the vector of Pauli matrices. 

For temperatures much lower than the Fermi energy, assuming an isotropic momentum relaxation time $\tau_p$, the linearized diffusive transport equations for two dimensional charge and spin densities $\delta n$ and $\bf S$, respectively have the following form~\cite{Bernevig,BurkovPRB2004,note1}:
\begin{eqnarray}
\nonumber
\dot{\delta n} &=& D \partial_k (\partial_k \delta n - \nu e E_k) + 	2 \Gamma \epsilon_{ik}\partial_i {\tilde S}_k
\\
\nonumber
{\dot S}_i &=& D (\partial_k)^2 {\tilde S}_i + \sqrt{\frac{4D}{\tau_s}} \partial_i {\tilde S}_z - \frac{{\tilde S}_i}{\tau_s} 
	- \frac{\Gamma}{2} \epsilon_{ik} (\partial_k \delta n - \nu e E_k) 
\\
\label{vector_equations}
{\dot S}_z & = &  D (\partial_k)^2 {\tilde S}_z 
 	- \sqrt{\frac{4D}{\tau_s}} \partial_k {\tilde S}_k 
	- \frac{2 {\tilde S}_z}{\tau_s} 
\end{eqnarray}
Here $i,k \in \{ x, y \}$, $\nu = m/(\pi \hbar^2)$ is the density of states, and $D = v_F^2\tau_p/2$  is the diffusion coefficient. The parameter $\Gamma = 2 \alpha (\epsilon_{\alpha} \tau_p)^2/(1 + 4 (\epsilon_{\alpha} \tau_p)^2)$ and the spin relaxation time $\tau_s = \tau_p (1 + 4 (\epsilon_{\alpha} \tau_p)^2)/2 (\epsilon_{\alpha} \tau_p)^2$ are expressed through the spin-orbit splitting $\epsilon_{\alpha}  = \alpha p_F$ , where $v_F$ and $p_F$ are the Fermi velocity and wave vector, respectively \cite{MishchenkoPRL2004}. The renormalization ${\tilde {\bf S}} = {\bf S} - \nu g \mu_B {\bf B}$, where $g$ is the Lande's g factor, accounts for the Zeeman action of the magnetic field. It is only significant in the presence of a large externally applied magnetic field and can be neglected here.

The electric field ${\bf E} = {\bf E}_{ext} + {\bf E}_{ind}$ in Eq.~(\ref{vector_equations}) is a sum of the externally applied field driving the system, and the field induced by charges and currents in the 2DEG. First, we explore the effect of the induced electric field on the spectrum of the propagating mode. The speed of this mode is much smaller than the speed of light~\cite{Bernevig}, hence the induced field is quasistatic: ${\bf E}_{ind} = -\nabla \phi_{ind} + \dot{\bf A}_{ind}/c$ with potentials determined by the instantaneous values of electric charges and currents. The first term describes the Coulomb repulsion of induced charges and the second term accounts for the self-induction of the induced currents. The latter is small compared to the former, and will be neglected. As a result
\begin{equation}
\label{kernel:def}
E^{ind}_i({\bf r}) = \int {\mathcal F}_i({\bf r}-{\bf r}') e \delta n({\bf r}')  d^2 r' \ ,
\end{equation}
where ${\bf r}$ is the in-plane coordinate and the form of the Coulomb kernel ${\mathcal F}_i$ depends on the geometry of the experiment and the presence or absence of the gate electrode, as will be discussed below. 

\begin{figure}[b]
    \resizebox{.4\textwidth}{!}{\includegraphics{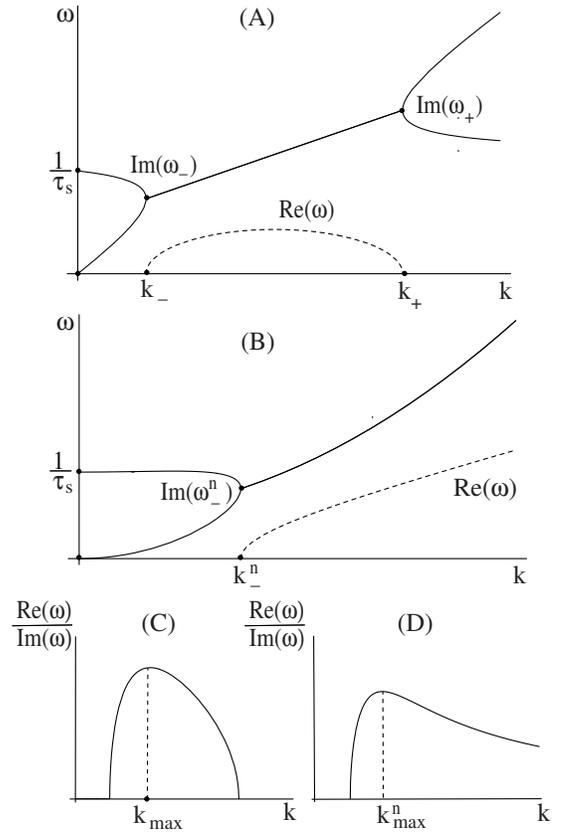}}
\caption{Schematic plots of the real (dashed line) and imaginary (solid line) parts of $\omega_{1,2}(k)$. A - with Coulomb repulsion, B - neutral carriers, C - ratio $Re(\omega)/Im(\omega)$ corresponding to plot A, D - same ratio for plot B.}
 \label{fig:sketch_re_im}
\end{figure}

The propagating mode is a periodic solution of Eqs.~(\ref{vector_equations}) with a density component $\delta n e^{ikx -i\omega t}$ and a single nonzero spin component $S_y(k,\omega) e^{ikx -i\omega t}$ (we choose $\hat x || {\bf k}$) \cite{Bernevig}. The induced electric field of this mode is directed along $\bf k$. Introducing the spin-diffusion length $l_s = \sqrt{D \tau_s}$, and defining dimensionless quantities ${q} = k l_s$, ${q}_B = \nu e^2 l_s$, and $\gamma = \Gamma \sqrt{\tau_s/D}$, we perform Fourier transforms on Eqs.~(\ref{vector_equations}) with the result,
\begin{eqnarray}
\nonumber
&& \left|
	\begin{array}{cc}
		i\omega - \phi_1/\tau_s & 2 i \gamma q/\tau_s
		\\
		i \gamma \phi_1/2q\tau_s & i\omega - \phi_2/\tau_s
	\end{array}
\right| 
\left( \begin{array}{c} \delta n \\ S_y \end{array} \right) =
\left( \begin{array}{c} i q \\ \gamma \end{array} \right)
\frac{\sigma_{0}}{e l_s} E^{ext}_x
\\
\label{fourier}
&&	
	\phi_1({q}) = {q}^2 \left( 1 + iF\frac{{q}_B}{{q}} \right)
	, \quad
	\phi_2({q}) = {q}^2 +1 \ ,
\end{eqnarray} 
where $F$ is the Fourier transform of the kernel of Eq.~(\ref{kernel:def}) $F(k) = \int {\mathcal F}_x({\bf r}) e^{-i k x} dx dy$, and $\sigma_0 = \nu e^2 D$. The spectrum of the mode is then given by
\begin{equation}
\label{eigenmodes}
\omega_{1,2} = \left(\frac{1}{\tau_s}\right) \frac{
-i (\phi_1 + \phi_2) \pm \sqrt{4 \gamma^2 \phi_1 - (\phi_1 - \phi_2)^2}
	}{2}
\end{equation}
Coulomb interaction enters the spectrum through the renormalization factor $R = 1 + i F q_B/q$ with large parameter
 ${q}_B \gg 1$, reflecting the strength of the Coulomb repulsion. For the case considered in Ref.~\onlinecite{Bernevig}, $D = 10$cm$^2$/s, $\tau_s = 1$ ns, $\tau_p = 1$ ps, $m/m_e = 0.07$ we estimate ${q}_B \approx 450$.

For small wave vectors both $\omega_{1,2}(k)$ are imaginary, corresponding to two purely dissipative modes of the system: the charge diffusion and the spin relaxation. As $k$ is increased, $\omega_{1,2}(k)$ acquire a real part when the expression under the square root in Eq.~(\ref{eigenmodes}) becomes positive, and the mode can become weakly damped if the condition $Re(\omega)/Im(\omega) \gg 1$ is satisfied \cite{Bernevig}. The existence of such a regime depends on the value of $\gamma$, and the properties of the function $F(k)$. First, consider a single 2DEG layer of thickness $d_z$. Since the diffusive transport equations are only valid for large wavelength $2\pi/k \gg l$, where $l = v_F \tau_p \gg d_z$ is the electron mean free path, we can expand $F(k)$ in powers of small $k d_z$ and obtain:
$$
F_s(k) = - 2 \pi i + \ldots
$$
Second, consider a 2DEG with a gate electrode located at a distance $d_g > d_z$. Here the kernel becomes smaller due to partial screening by image charges. Expansion in $k d_z, k d_g \ll 1$ gives
$$
F_g(k) = -2\pi i \left( d_g k - \frac{d_z k}{3} \right) + \ldots
$$
Sketches of the $\omega_{1,2}(k)$ functions are shown in Fig.~\ref{fig:sketch_re_im}. The spectra in the single-layer and gated cases have the same overall shape. Due to Coulomb repulsion, the real part of $\omega_{1,2}(k)$ exists in a finite interval $[k_{-},k_{+}]$, in contrast to the case of neutral carriers where it exists for $k > k_{-}^n$. The mode propagates with least dissipation for the wave vector $k_{\max}$ at which $|Re(\omega)/Im(\omega)|$ reaches a maximum. One can prove that, in the physically relevant regime, ${q}_B \gg \gamma$, the Coulomb renormalization in a single layer 2DEG, $R_s = 1 + 2\pi q_B/q \approx 2\pi q_B/q \gg 1$, is large in the relevant interval of wave vectors. In the gated 2DEG $R_g = 1 + 2\pi{q}_B (d_g - d_z/3)/l_s \gg 1$ also holds for typical systems. In the limit of large Coulomb renormalization, our calculation gives the following values for the characteristic parameters of the curves shown in Fig.~\ref{fig:sketch_re_im}(A).
\begin{eqnarray}
\nonumber
&& k_{\pm}^s = \frac{(\sqrt{\gamma^2 + 1} \pm \gamma)^2}{2\pi{q}_B l_s},
	\quad k_{\max}^s = \frac{1}{2\pi{q}_B l_s}
\\
\nonumber
&& k_{\pm}^g = \frac{\sqrt{\gamma^2 + 1} \pm \gamma}{l_s \sqrt{R_g}} ,
	\quad k_{\max}^g = \frac{1}{l_s \sqrt{R_g}}
\\
\label{s_params}
&& Im(\omega_{\pm}) = \frac{\gamma^2 + 1 \pm \gamma\sqrt{\gamma^2 + 1}}{\tau_s}
\\
\nonumber
&& Re(\omega_{\max}) = \frac{\gamma}{\tau_s},
	\quad 
	\max\left|\frac{Re(\omega)}{Im(\omega)}\right| = \gamma 
\end{eqnarray}
For neutral carriers the characteristic parameters (see Fig.~\ref{fig:sketch_re_im}(B)) are 
\begin{eqnarray}
\nonumber
&& k_{-}^n = \frac{1}{2\gamma l_s}, 
	\quad k_{\max}^n = \frac{\sqrt{\gamma^2 + 1}}{\gamma l_s \sqrt{2}} 
\\
\label{n_params}
&& Im\left(\omega_{-}^n \right) = \frac{2\gamma^2 + 1}{2\gamma^2 \tau_s},
\\
\nonumber
&& Re(\omega_{\max}^n ) =  \frac{\sqrt{2\gamma^2 + 1}}{2\tau_s}, \quad
\max\left|\frac{Re(\omega)}{Im(\omega)}\right| = \frac{\gamma^2}{\sqrt{2\gamma^2 + 1}}
\end{eqnarray}
Eqs.~(\ref{s_params}) and (\ref{n_params}) show that the Coulomb interaction increases the wavelength of the propagating mode. At the point of least damping the wavelengths are $2\pi/k^s_{\max} = 4\pi^2{q}_B l_s$ and $2\pi/k^g_{\max}= 2\pi l_s \sqrt{R_g}$ in the single and gated cases, respectively. Both wavelengths are much larger than for the neutral carrier case, $2\pi/k_{\max}^n \sim l_s$. The increased wavelength allows the system to reduce the Coulomb repulsion. Accordingly, a smaller increase is observed in the partially screened gated case ($\sqrt{R} < 2\pi{q}_B$). Still, in both cases the wavelength can become comparable to the size of a typical 2DEG sample, at which point the boundary conditions, not considered here, will start to play an important role. The condition for the mode to be weakly damped, however, is unchanged by the Coulomb repulsion and remains $\gamma \gg 1$ \cite{caution}. Note that the diffusive regime condition, $2\pi/k \gg l$, is well satisfied in the large $R$ case.

\begin{figure}[t]
    \resizebox{.35\textwidth}{!}{\includegraphics{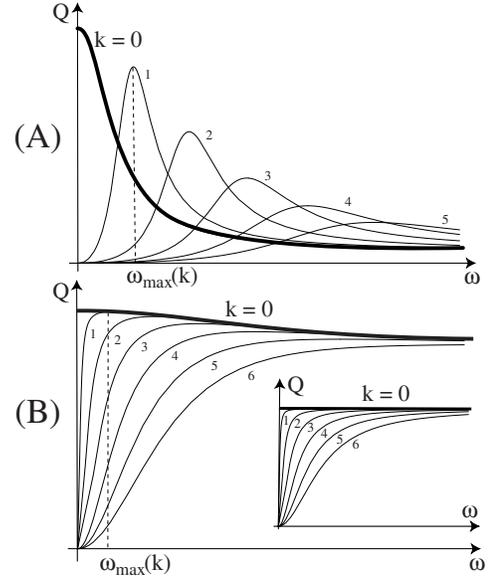}}
\caption{Numbered lines represent absorption curves for representative wave vectors $k_1 < k_2 < \ldots < k_6$. Thick solid lines correspond to $k=0$. As $k$ is increased, the position of the maximum of absorption moves to higher frequency until the maximum disappears. A - case of $\gamma > 1$, B - case of $\gamma < 1$. Inset of B - case of $\gamma = 0$.}
 \label{fig:absorption}
\end{figure}

A spatially periodic external electric field can excite the propagating mode and one expects that in the weakly damped regime a maximum of energy absorption at the frequency $\omega = Re(\omega_{1,2}(k))$ will  be a signature of the mode's presence. Experimentally, the spatial periodicity of the driving field can be created by an array of gates, as in experiments on plasmon resonance \cite{allen_plasmon}, or, optically, using crossed light beams. The average dissipated power equals $Q = E_{ext}Re(j)/2$, where the Fourier component of the current $j(k,\omega)$ can be found from the continuity equation $e \delta\dot n + {\rm div} j = 0$ as $j = e (\omega/k)\delta n$. Finding $\delta n$ from Eq.~(\ref{fourier}) one obtains:
\begin{eqnarray}
\label{absorption}
Q(w) &=& \left( \frac{w^2(w^2 + A)}{w^4 + w^2 B + C^2} \right) \frac{\sigma_{0} E_{ext}^2}{2}
\\
\nonumber
A &=& \phi_2^2 + \gamma^2(\phi_1 + 2\phi_2)), \quad
B = \phi_1^2 + \phi_2^2 - 2\gamma^2\phi_1
\\
\nonumber
C &=& \phi_1 (\phi_2 + \gamma^2)
\end{eqnarray}
Representative absorption curves are shown in Fig.~\ref{fig:absorption}. For $\gamma > 1$ the peak is indeed observed for $k$ values supporting the propagating mode. However, $Q(\omega)$ also exhibits a maximum for small wave vectors where $\omega_{1,2}(k)$ are purely imaginary. The maximum is present even in the $\gamma \ll 1$ case. The exact condition for the existence of the maximum is $A>B$ or, equivalently, $k < k_*$, where $k_*$ is determined by the equation $\gamma^2(3\phi_1(k_*) + \phi_2(k_*)) = \phi_1(k_*)^2$. All maxima exist in the diffusive regime, $2\pi/k_* \gg l$.

To understand the presence of an absorption maximum in a system with two purely diffusive modes, note that the spin-orbit interaction renormalizes the DC conductivity to $\sigma = (1 + 2\gamma^2)\sigma_{0}$. At finite frequencies the conductivity decreases and reaches its bare value $\sigma_0$ at $1/\tau_s \ll \omega \ll 1/\tau_p$. Accordingly, the $k=0$ curves in Fig.~\ref{fig:absorption} are bell-shaped with a maximum at $\omega = 0$. But at any $k > 0$ the DC current is required to vanish. As a result, for $k \neq 0$ the curve $Q(\omega)$ starts from zero and then rapidly approaches the $Q(\omega)|_{k=0}$ curve, thus creating a pronouncedly asymmetric maximum, which looks more like a dip at $\omega < \omega_{\max}(k)$ (see Fig.~\ref{fig:absorption}(B)). In the absence of spin-orbit interaction the $k=0$ curve becomes a horizontal line (inset in Fig.~\ref{fig:absorption}(B)). Thus an observation of the bell-shaped $Q(\omega)|_{k=0}$ curve can serve as a proof of sufficient experimental sensitivity, allowing detection of the spin-orbit interaction.

The maxima observed at $\gamma > 1$ are much more symmetric and significantly extend above the $Q(\omega)|_{k=0}$ curve as shown in Fig.~\ref{fig:absorption}(A). Notably, the shape of $Q(\omega)$ continuously evolves with $k$, showing no singularities at the boundaries $k_{\pm}$ of the interval of the propagating mode's existence. The distinction between the systems with and without the propagating mode has to be made by matching the experimentally measured $Q(\omega,k)$ familiy of curves to Figs.~\ref{fig:absorption}(A) or~\ref{fig:absorption}(B).

To conclude, the propagating mode can be viewed as a periodic involvement of direct and inverse spin-galvanic effects and thus called a ``spin-galvanic'' mode. In this mode the spin-orbit interaction plays the roles of both inertia and a restoring force. Oscillations occur between two states with the same spin-orbit energy: one with non-equilibrium spin density and charge current, and the other with non-equilibrium spatial charges and spin current. The frequency of the mode is of the order $\omega \lesssim 1/\tau_s \ll 1/\tau_p$. The spin-galvanic mode has no connection with the two-dimensional plasmon modes~\cite{litovchenko_plasmon,theis_plasmon}. Although the spectrum of these plasmons $\omega \sim \sqrt k$ does not have a gap, they become damped and unobservable for $\omega < 1/\tau_p$. The inertial element of plasmon oscillations is the mechanical inertia of electrons, thus they cannot survive in the diffusive regime.

We have shown that the Coulomb repulsion of electrons modifies the properties of the mode, but does not suppress it. The low-frequency spin-orbit mode still exists for $\Gamma \gg \sqrt{D/\tau_s}$, similar to the way the low frequency modes exist in magnetic field~\cite{theory_of_metals}. However, the Coulomb interaction modifies mode's spectrum and significantly increases its wavelength both in single layer and gated 2DEGs. This has important implications for the experimental observation of the mode.  On the one hand, spatially periodic electric field with larger wavelength can be created by purely optical methods, without the need to deposit periodic arrays of gate electrodes. On the other hand, it has to be checked that the wavelength of the mode does not approach the size of the sample. Theoretical description of the latter situation will require a derivation of appropriate boundary conditions and a recalculation of the resonant frequency.

The original paper \cite{Bernevig} suggested that a narrow charge packet can be used to probe all wave vectors in the interval of the mode's existence. The evolution of such a packet would reflect the presence or the absence of the mode. Since an experimental method of injecting the excess charge into a 2DEG is not obvious, and since the domain of the spin-galvanic mode's existence is narrowed to a finite interval of wave vectors by the Coulomb interaction, it may be beneficial to probe the mode at a given wave vector, as suggested in this Letter. We have shown that such an experiment is capable of conclusively proving either the presence or the absence of the mode by observing the frequency dependence of the absorption of high frequency electromagnetic field, spatially periodic in the plane of 2DEG.

The authors are grateful to Dan Arovas, Andrei Bernevig, Rai Moriya, Emmanuel Rashba, Glen Solomon, and Shou-Cheng Zhang for illuminating discussions. This research was supported by DMEA contract \# H94003-04-2-0404.


\begin{thebibliography}{2}
\bibitem{caution} 
Condition $\gamma \gg 1$ is incompatible with a usual Dyakonov-Perel model, where $\gamma \leq \alpha/2v_F \ll 1$ holds. This limitation and ways to overcome it are discussed in Ref.~\onlinecite{Bernevig}.
\end{thebibliography}
\end{document}